\documentclass{article}
\usepackage[margin=1in]{geometry}
\usepackage[utf8]{inputenc}
\usepackage{todonotes}
\usepackage[colorlinks,allcolors=cyan!70!black]{hyperref}
\usepackage{amsmath}
\usepackage{amsfonts}
\usepackage{subfigure}
\usepackage{graphicx}
\usepackage{lipsum}
\usepackage{multicol}

\title{Applying an electrostatic cross-correlation to the CFTR-ATP interaction}
\author{Alex Saad-Falcon, Mark Bolding, James Dee, Ryan S. Westafer, and Douglas R. Denison\footnote{Georgia Tech Research Institute (GTRI)} \\
Nael McCarty\footnote{Emory University, Emory+Children's Cystic Fibrosis Center of Excellence} \\
William D. Hunt\footnote{Georgia Tech School of Electrical and Computer Engineering}}
\date{May 2021 (Updated April 2024)}

\begin{document}
\newcommand{\vr}{\mathbf{r}}
\newcommand{\rplus}[2]{\ref{#1}\hyperref[#1]{#2}}

\maketitle



\begin{abstract}
The cystic fibrosis transmembrane conductance regulator (CFTR) is an important membrane protein in vertebrates. The function of CFTR is to transport chloride ions across the cell membrane, which is known to require adenosine triphosphate (ATP). Whereas most conventional wisdom suggests that ATP interacts with CFTR purely through random collisions via diffusion, we investigate electrostatic interactions between CFTR and ATP at the mesoscale (10s of Angstroms). We use molecular dynamics to simulate CFTR-ATP interactions in cases where CFTR is bound/unbound from ATP, and we demonstrate an electrostatic potential gradient towards CFTR when ATP is unbound. We additionally compute electrostatic interactions between ATP and the solvent and membrane, which are simulated explicitly.

\end{abstract}
\section{Introduction}


The cystic fibrosis transmembrane conductance regulator (CFTR) is a chloride ion channel whose dysfunction causes cystic fibrosis. CFTR is fueled by adenosine triphosphate (ATP), a common energy source in cells. The objective of this work is to elucidate the receptor-ligand interaction between CFTR and ATP. Currently, the path that an ATP molecule takes when binding to a CFTR protein is unknown. We investigate if the trajectory of ATP binding to CFTR is mediated at distances larger than the contact regime. At these intermolecular distances in water, only the Coulomb force has a strength comparable to random thermal forces. Therefore, the focus of this work is on the electrostatic interactions of ligand and receptor before binding. We also compare the receptor-ligand interaction with the interaction between the ligand and other system components, such as solvent and cell membrane.

To investigate these statistical interactions, we use molecular dynamics with multiple random initializations to simulate the CFTR protein and its interactions with an ATP molecule in solution at multiple ranges. We then apply an electrostatic correlation called the molecular ambiguity function (MAF) \cite{maf} to evaluate the ATP-protein interactions over each trajectory. Additionally, we use the static conformations of CFTR and ATP at each frame and evaluate the MAF over a grid of ATP translations to discover interesting contours.

Electrostatic correlations as applied to analyzing molecular dynamics simulations are not novel. The electrostatic force is just one constituent of the several forces included in molecular dynamics additive force fields. Other names referring to the same concept are electrostatic potential and electrostatic free energy. However, when evaluating the electrostatic potential, the interaction between each atom and every other atom is utilized. The molecular ambiguity function, on the other hand, evaluates the cross-correlation between two groups of atoms, and it does not include self-interactions between atoms in each of the groups.

There is a large body of work in molecular docking on the basis of shape and electrostatic complementarity at close range \cite{katchalski-katzir_molecular_1992,gabb_modelling_1997}, and 
some works have shown that long range interactions may be cancelled out by solvent forces \cite{gumbart_efficient_2013}.
In contrast, we investigate the global electrostatic landscape at ``long range,'' requiring modeling of the whole system before the ligand reaches the neighborhood of the binding site. Accurate representation of the full CFTR structure supports modeling of the longest range interactions.

\section{CFTR Structure and Function}

CFTR is a complex polytopic membrane protein that is the locus of the primary defect in the lethal genetic disease: cystic fibrosis.  A member of the ATP-binding cassette (ABC) transporter superfamily of proteins, it is unique in functioning as an ATP-gated ion channel \cite{csanady_structure_2019}. Hence, the molecular evolution of function in CFTR built upon the large conformational changes that its closest relatives undergo in the process of active transport of substrates across the membrane and converted those conformational changes into the establishment of an open pore across the membrane, enabling passive transport of chloride ions down the electrochemical gradient \cite{srikant_evolutionary_2020}. This provides extraordinary opportunities to model ion channel function, and to understand the role of individual channels on cell physiology. 

Because members of this superfamily undergo large conformational changes during their functional cycle as ATP-driven transporters, CFTR also undergoes large conformational changes during its function as an ATP-gated channel.  At the same time, because CFTR's function leads to the movement of charged ions across the membrane, i.e. an electrical current, CFTR function can be studied with high resolution using the patch clamp technique \cite{cui_electrophysiological_2021}.  

With the recent publication of several high resolution structures of CFTR in a variety of conditions, we now can also apply molecular dynamics approaches to understand CFTR function. We can ask how channel conformation relates to channel function, and can formulate hypotheses regarding the effects of external forces (such as electric fields) on channel structure and test them via experimental assays.  

Finally, even though ABC transporters have been under intense study for many years due to their roles in a wide array of cellular functions, fundamental questions remain regarding their function including understanding how ATP accesses the binding sites at the interfaces of the nucleotide binding domains (NBDs), and how adenosine diphosphate (ADP) - the product of ATP hydrolysis - exits the binding sites at the end of the gating cycle.  By linking structural analysis of CFTR by molecular dynamics with functional analysis by patch clamp, and with the addition of newly developed methods such as the molecular ambiguity function (MAF) \cite{maf}, we are now positioned to answer important questions regarding these clinically relevant proteins.

There are many open questions pertaining to the function and dysfunction of CFTR. For example, the pathways ATP takes to arrive at the consensus binding sites (internal to CFTR) are unknown. However, anion pathways are known, and an ATP reentry pathway has been postulated to exist and involve ``separation of the NBD dimer interface around site 2 uncoupled from pore closure'' \cite{csanady_structure_2019}.
Another open question is whether the interior state of the protein influences the external free energy landscape (MAF), which may alter affinity for ATP or its path to the (competent) binding site.
We have employed all-atom molecular dynamics simulations with long-range electrostatics in attempt to better understand the likely interactions and/or pathways for ATP in the vicinity of CFTR.




\section{Methods}
\subsection{Molecular dynamics} \label{sec:md_methods}

In this work, all molecular dynamics simulations were performed using the NAMD/VMD suite \cite{namd}. The CFTR protein was acquired from the Research Collaboratory for Structural Bioinformatics (RCSB) Protein Data Bank (PDB) \cite{pdb} using the PDB identifier 6MSM \cite{6msm}. This identifier is a cryo-electron microscopy (Cryo-EM) model structure of a mutated version of human CFTR with ATP bound at both the active and inactive binding sites. The E1371Q mutation in 6MSM precludes ATP hydrolysis at the competent binding site (ABS2), which prolongs the lifetime of the open CFTR channel and enables the production of the Cryo-EM structure for ATP-bound nearly-open-channel CFTR. 

Also appearing in the 6MSM structure are several loose hydrocarbons of varying lengths (nearby tails from phospholipids) and a mediating cholesterol molecule along one of the transmembrane domains. Additionally, one magnesium ion was found in each of the CFTR binding sites. The cholesterol molecule was preserved for simulation, since membrane cholesterol has been shown to affect CFTR ion channel activation \cite{cftr_chol}. The magnesium ions in the binding sites were also preserved, and the incomplete hydrocarbon tails were discarded to later be replaced with full phospholipids. 

To further prepare the structure for simulation, the E1371Q mutation was reverted to create wild-type CFTR. The resulting structure was aligned and placed in 125~\AA$\times$125~\AA~of POPC membrane using VMD. The purpose of such a large membrane is to ensure the inter-channel spacing (via periodic boundary conditions) is large enough that inter-channel effects can be ignored, and to provide room for the new, unbound ATP that will be introduced. Any POPC molecules within close proximity of the protein or cholesterol were removed. This system was then solvated and ionized with 150~mM NaCl.

Two experiments with ten random initializations each were then conducted using this system. For one, the ATP at the active ATP binding site was removed, and for the other, it was left alone. For both experiments, a new ATP (the ``free ATP") was introduced in solvent around 10~\AA~away from the intracellular side of the protein with a random orientation. For each of the ten runs in each experiment, the system was ionized independently so as to avoid any bias from initial ion placements. All of the twenty total runs were minimized/equilibrated in stages starting with just the solvent, then adding the membrane, and then adding the protein and free ATP with restraints. These equilibration stages lasted a total of 2.5~ns. Given the sudden removal of the bound ATP for the unbound experiments, a slow equilibration was required, and the equilibration was verified to not significantly disrupt the CFTR structure.

\begin{table}[h]
\label{table:md}
\begin{center}
\begin{tabular}{ll}
\textbf{Setting}          & \textbf{Value} \\
Parameters                & CHARMM36~\cite{charmm} \\
Number of atoms           & 1.1~million \\
Timestep                  & 1~fs \\
Equilibration Temperature & 300~K \\
Production Temperature    & 310~K \\
Equilibration Time        & 2.5~ns\\
Production Time           & 10~ns \\
Langevin Damping          & 1/ps \\
Langevin Piston           & 1~atm \\
Langevin Piston Period    & 200~fs \\
Langevin Piston Decay     & 50~fs \\
Cutoff                    & 12~\AA \\
PME                       & Yes \\
Steps per Cycle           & 10
\end{tabular}
\caption{A summary of parameters and values used for molecular dynamics.}
\end{center}
\end{table}

\begin{figure}
\centering
\subfigure[The initial simulation configuration. The free ATP molecule is visible at the top.]{\label{fig:cftr_maf_system:a}\includegraphics[width=0.45\textwidth]{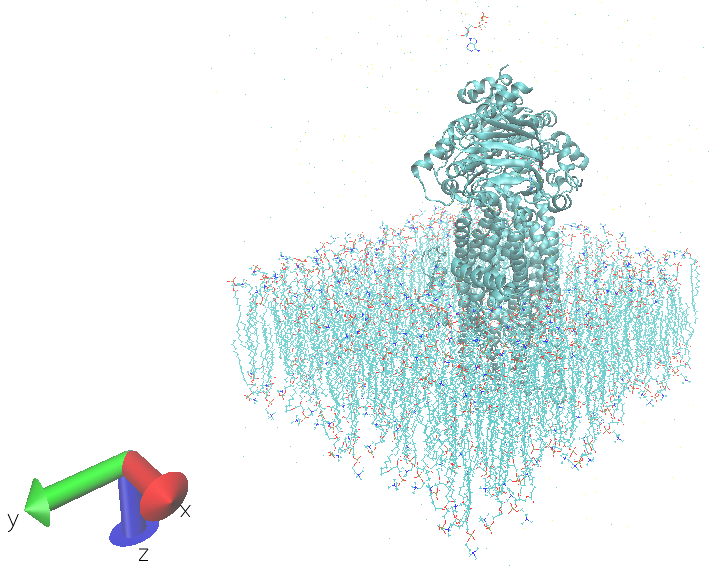}}
\subfigure[A system configuration during production for one of the runs. The free ATP molecule is visible at the top and left of center.]{\label{fig:cftr_maf_system:b}\includegraphics[width=0.45\textwidth]{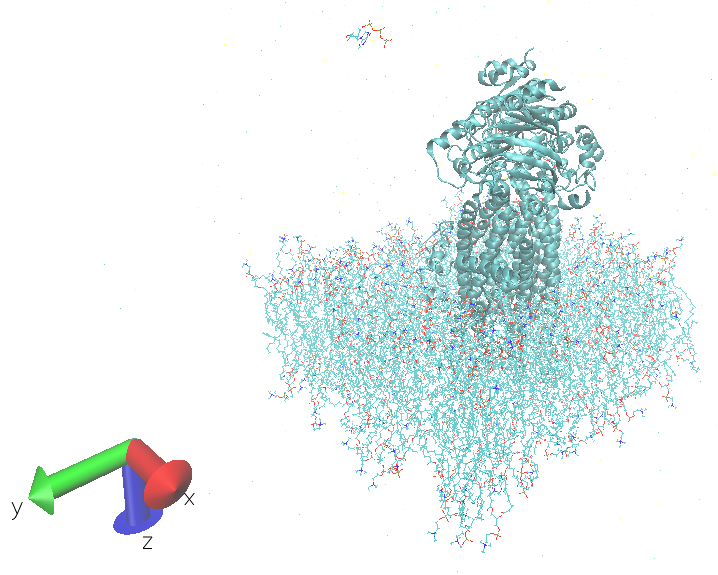}}
\caption{Snapshots of the system configuration at initialization and during production. Solvent molecules are not shown.}
\label{fig:cftr_maf_system}
\end{figure}

Finally, ten production runs were simulated for each experiment. The free ATP was released at the start of production, and the different equilibration simulations for each run ensured different initial conformations for the protein and free ATP at the beginning of production. Over the course of each run, trajectories were saved every 1~ps, and these trajectories are the primary drivers of the results in Section \ref{sec:results}. A summary of the molecular dynamics parameters used is shown in Table \ref{table:md}. The initial system and a later system for one of the runs are shown in Figure \ref{fig:cftr_maf_system}.

\subsection{Molecular Ambiguity Function}
In signal processing the ambiguity function $\xi$ expresses the response of a filter to an input signal. It is defined to be the response of the filter as a function of two parameters which represent time and phase delays applied to the filtered signal. Mathematically the ambiguity function is given by the equation
\begin{align*}
    \xi(\tau, f) = \int s(t) s(t-\tau) e^{i 2 \pi f t} dt.
\end{align*}
The indicated integration has limits $\tau \in (-d_\tau, d_\tau)$, $f \in (-d_f, d_f)$ where $d_\tau$ and $d_f$ are chosen based on the specific filter and situation under consideration.

In analogy with the ambiguity function, the MAF expresses the electric potential interaction between molecules as a function of system configuration. As with the ambiguity function, a small set of parameters are chosen to represent the system configuration. These may be the relative position and orientation of the molecules, for example, or the elapsed time in a molecular dynamics simulation. Since it is the second of these that interests us here, the molecular ambiguity function $M$ is defined to be
\begin{align} \label{maf_def}
    M(t) = \epsilon_0 \int E_r(t, \mathbf{x}) \cdot E_l(t, \mathbf{x}) d\mathbf{x}
\end{align}
where $E_r$ is the electric field of the receptor molecule (which may include solvent and additional molecules) and $E_l$ is the electric field of the ligand, in this case ATP. The parameters $t$ and $\mathbf{x}$ indicate simulation time and spatial position, respectively. The indicated integral is over all space. MD simulation time accounts for different conformations of the receptor and ligand, since the positions of atoms are not fixed relative to each other.

By modeling the electric field as coming from collections of point particles, the MAF can be written in a computationally favorable way as
\begin{align} \label{maf_eval}
    M(t) = k_e \sum_{i=1}^{m_1} \sum_{j=1}^{m_2} \frac{q_i q_j}{r_{ij}(t)}.
\end{align}
The number of atoms in molecule 1 is $m_1$ and the number of atoms in molecule 2 is $m_2$. $r_{ij}(t)$ is the distance between atom $i$ of molecule 1 and atom $j$ of molecule 2 at simulation time $t$. Similarly $q_i$ and $q_j$ are corresponding atomic charges. With this formulation the capabilities of available hardware allow for a dramatic increase in the speed of MAF computations. 
For systems including millions of atoms in the receptor, the computation finishes in about one second on a desktop CPU.

Since the electric field is proportional to the gradient of the potential, it is intuitively reasonable that the integral (\ref{maf_def}) results in the sum of electrostatic potentials that appears in (\ref{maf_eval}). In NAMD the total potential can be separated into two terms, one consisting of interactions due to bonded atoms (denoted $U_b$) and the other of interactions between non-bonded interactions (denoted $U_{nb}$) \cite{namd_potentials}. Since the goal of this work is to predict long range interactions between molecules as depicted by MD simulations, it is reasonable to ask what the effect of the other potentials is at long range.

The potential between bonded atoms affects the internal configuration of the atoms in each molecule. MD simulations account for these effects. $U_{nb}$ can be separated into two terms, one being the electrostatic potential and the other the Lennard-Jones potential $U_{lj}$. Since
\begin{align*}
    U_{lj} \propto \left( \frac{R}{r_{ij}} \right)^{12}-2 \left( \frac{R}{r_{ij}} \right)^6
\end{align*}
for a constant $R$ that depends on the pair of atoms considered, one can see that this interaction is weak for large values of $r_{ij}$. This leaves the electrostatic potential as the dominant contributor to $U_{nb}$.


\section{Results} \label{sec:results}

As described in Section \ref{sec:md_methods}, two experiments with random initializations were conducted with 10 production runs per experiment. Using the resulting trajectories, we computed various quantities including root mean square deviation (RMSD), root mean square fluctuation (RMSF), and molecular ambiguity function (MAF). 

\subsection{CFTR-ATP Interaction using MAF}
\begin{figure}
\centering
\subfigure[MAF for CFTR bulk-ATP with ATP in active site.]{\label{fig:mcmaf:a}\includegraphics[width=0.45\textwidth,trim={0.2cm 0.2cm 1.5cm 0.5cm},clip]{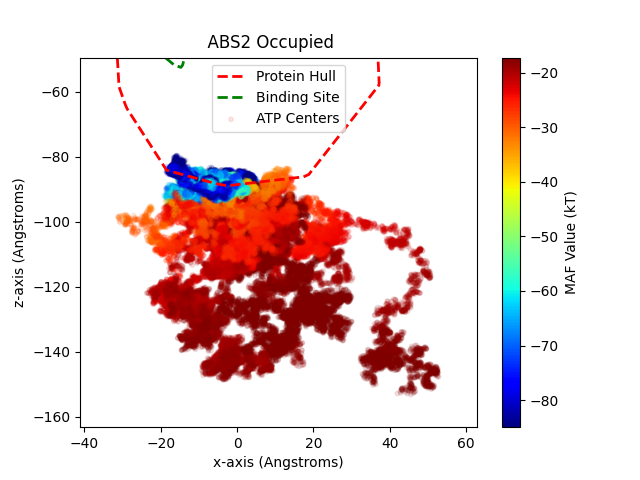}}
\subfigure[MAF for CFTR bulk-ATP without ATP in active site.]{\label{fig:mcmaf:b}\includegraphics[width=0.45\textwidth,trim={0.2cm 0.2cm 1.5cm 0.5cm},clip]{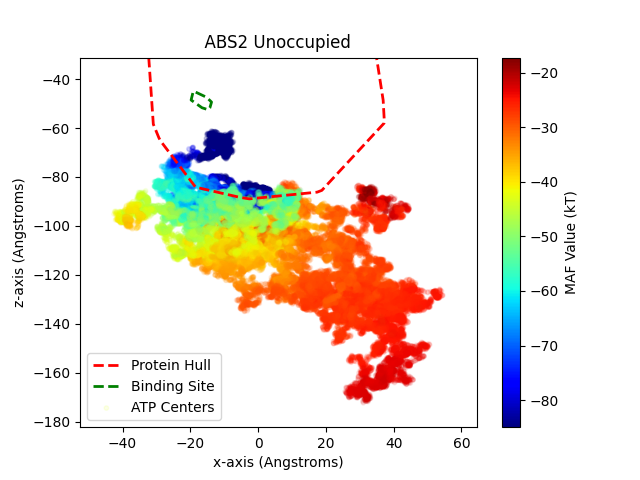}}
\caption{The molecular ambiguity function (MAF) for the CFTR bulk-ATP interactions over 10 trajectories for each of the two experiments. The colorbar minimum and maximum are chosen as the 10th and 90th percentiles for MAF values, respectively. A clear gradient towards the active binding site is observed in both experiments.}
\label{fig:mcmaf}
\end{figure}

The molecular ambiguity function (MAF) between the free ATP and the CFTR bulk over all 10 trajectories for each experiment is shown in Figure \ref{fig:mcmaf}. The CFTR bulk consists of the CFTR protein itself, the attending cholesterol, and the two magnesium ions at the binding sites. In each trajectory, the ATP explores different regions in the vicinity of the intracellular end of the CFTR. In Figure \ref{fig:mcmaf}{a}, one of the trajectories can be seen clearly as the ATP meanders in the negative z-direction. 

When aggregating the trajectories for the MAF of the protein versus the free ATP, a clear picture is seen - an electrostatic potential energy gradient generally decreasing toward the CFTR active binding site in the interior of the protein. Additionally, it is clear that the two experiments are on different levels energetically; the average MAF value for each experiment differs by around $20~kT$, as can be seen in the Figure \ref{fig:mcmaf} colorbars. 

This gradient exists both when the active binding site is already occupied by an ATP and when it is not. When the active binding site is occupied (Figure \ref{fig:mcmaf}{a}), the MAF tends towards $0~kT$ at further distances (around $x=20$~\AA), and hence will likely be blurred by thermal noise. However when the binding site is unoccupied (Figure \ref{fig:mcmaf}{b}), the MAF at these same distances is more favorable $-30~kT$ at larger distances. To fully contextualize this result, we need to investigate the ATP interaction with other system components.

\begin{figure}
\centering
\subfigure[MAF for solvent-ATP with ATP in active site.]{\label{fig:mcmaf_solvent:a}\includegraphics[width=0.45\textwidth,trim={0.2cm 0.2cm 1.5cm 0.5cm},clip]{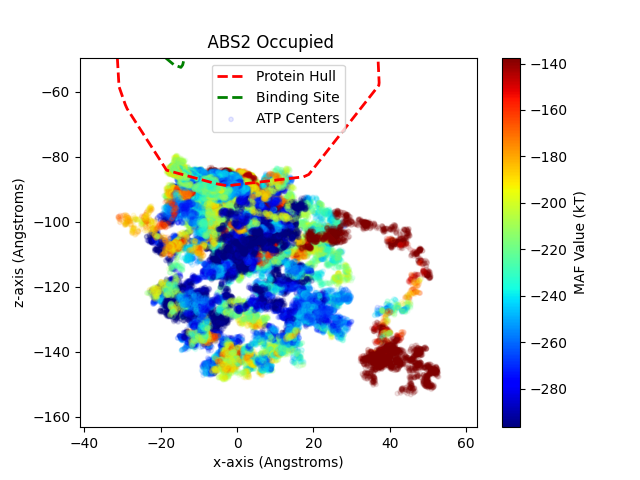}}
\subfigure[MAF for solvent-ATP without ATP in active site.]{\label{fig:mcmaf_solvent:b}\includegraphics[width=0.45\textwidth,trim={0.2cm 0.2cm 1.5cm 0.5cm},clip]{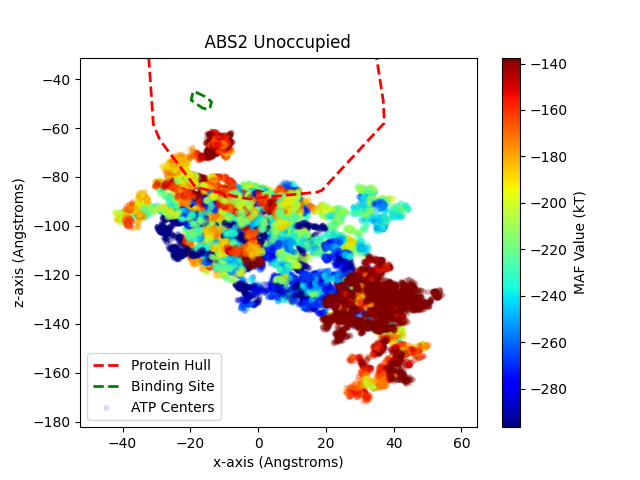}}
\caption{The molecular ambiguity function (MAF) for the solvent-ATP interactions over 10 trajectories for each of the two experiments. When compared to the CFTR-ATP interactions, these cover a greater dynamic range. }
\label{fig:mcmaf_solvent}
\end{figure}

\subsection{Solvent and Membrane Interactions}
The interaction between the CFTR protein and an external ATP does not occur in a vacuum --- this interaction happens in an intracellular environment with water, ions, and other cellular structures. The solvent in between the CFTR and ATP results in screening effects, where the inverse square scaling of the electric field with distance weakens into an exponential scaling. The solvent dipoles reorient to counteract the electric field, resulting in reduced electrical interactions except for at very short distances. The Debye length for water is taken to be 7-10~\AA \cite{debye}, though this typically only considers static fields at experimental timescales. It has been shown that under certain conditions, some frequencies and field strengths can surpass this limit \cite{maxwellfreq}.

At the time scale of molecular dynamics (fs to ns), the time it takes for water molecules and solvent ions to reorient in response to an electric field is not negligible. Because of this, it is desired to model solvent interactions either explicitly or implicitly. In order of increasing computational requirements, implicit solvent models include increasing the permittivity used in the MAF equation to that of water, applying an implicit solvent model such as the generalized Born implicit solvent, or estimating the effective medium between every pair of atoms in the MAF calculation. Explicit solvent interactions simply use the trajectories of water and ions directly from the molecular dynamics to estimate the resulting screening effects.

Because solvent trajectories were readily available from the molecular dynamics experiments, we computed the electrostatic correlation between the free ATP and all atoms of the solvent. This result is shown in Figure \ref{fig:mcmaf_solvent}. It is clear that solvent-ATP interactions are more noisy and less directed than CFTR-ATP interactions. Additionally, the solvent-ATP interactions are the primary energetic constituent of the net interaction between ATP and the molecular dynamics system. The solvent-ATP interactions form a deep electrostatic potential well which serves as a ``backdrop'' for other system interactions.

\begin{figure}
\centering
\subfigure[MAF for membrane-ATP with ATP in active site.]{\label{fig:mcmaf_membrane:a}\includegraphics[width=0.45\textwidth,trim={0.2cm 0.2cm 1.5cm 0.5cm},clip]{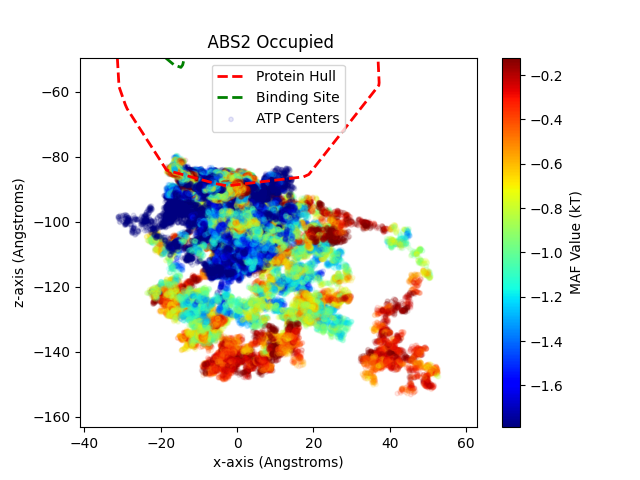}}
\subfigure[MAF for membrane-ATP without ATP in active site.]{\label{fig:mcmaf_membrane:b}\includegraphics[width=0.45\textwidth,trim={0.2cm 0.2cm 1.5cm 0.5cm},clip]{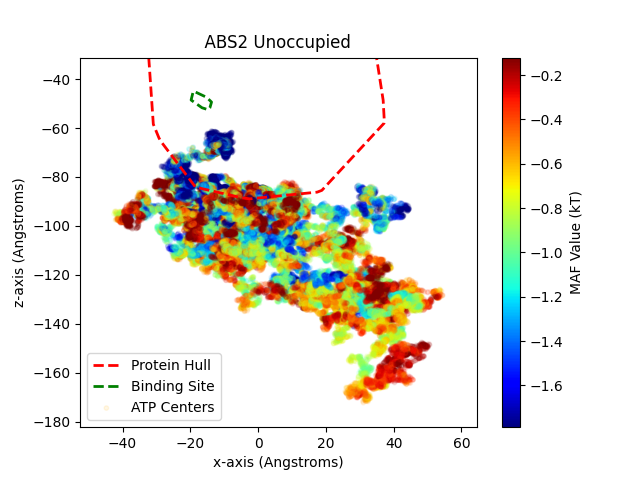}}
\caption{The molecular ambiguity function (MAF) for the membrane-ATP interactions over 10 trajectories for each of the two experiments. When compared to the CFTR-ATP and solvent-ATP interactions, these have a smaller dynamic range. }
\label{fig:mcmaf_membrane}
\end{figure}

The membrane-ATP correlation is shown in Figure \ref{fig:mcmaf_membrane}. This interaction is not as noisy as the solvent-ATP interaction, and it does not appear to have a preferential direction. In addition, the membrane-ATP interaction is of a much lower magnitude than the CFTR-ATP interaction because of the distance from the free ATP to the membrane, the membrane's zero net charge, and the membrane's relative immobility when compared to the solvent.

\begin{figure}
\centering
\subfigure[MAF for everything-ATP with ATP in active  site.]{\label{fig:mcmaf_else:a}\includegraphics[width=0.45\textwidth,trim={0.2cm 0.2cm 1.5cm 0.5cm},clip]{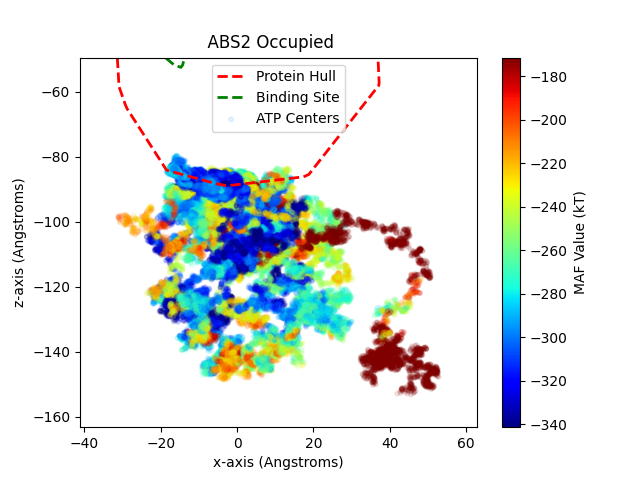}}
\subfigure[MAF for everything-ATP without ATP in active site.]{\label{fig:mcmaf_else:b}\includegraphics[width=0.45\textwidth,trim={0.2cm 0.2cm 1.5cm 0.5cm},clip]{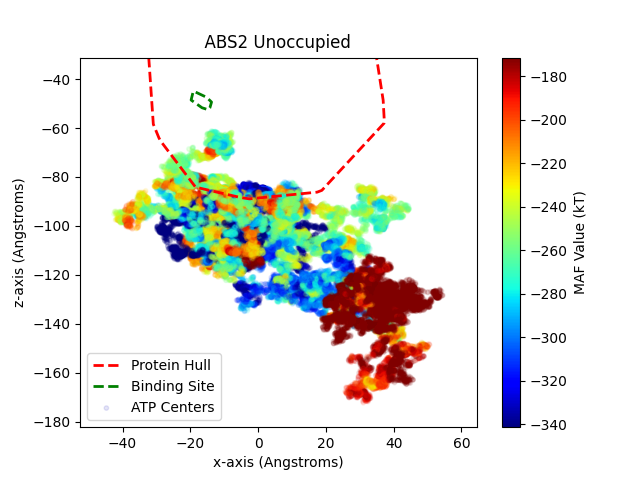}}
\caption{The normalized molecular ambiguity function (MAF) for the interaction between ATP and all other system components over 10 trajectories for each of the two experiments. This shows how the constituent components of the electrostatic interactions sum into the net interaction. }
\label{fig:mcmaf_else}
\end{figure}

When combined, these three interactions (CFTR bulk, solvent, and membrane) constitute the interaction of ATP with everything else in the molecular dynamics system. This results in the net system MAF shown in Figure \ref{fig:mcmaf_else}. While some gradient towards CFTR is seen, solvent interactions wash out much of the CFTR-ATP interaction.

\begin{figure}
\centering
\includegraphics[width=1.0\textwidth]{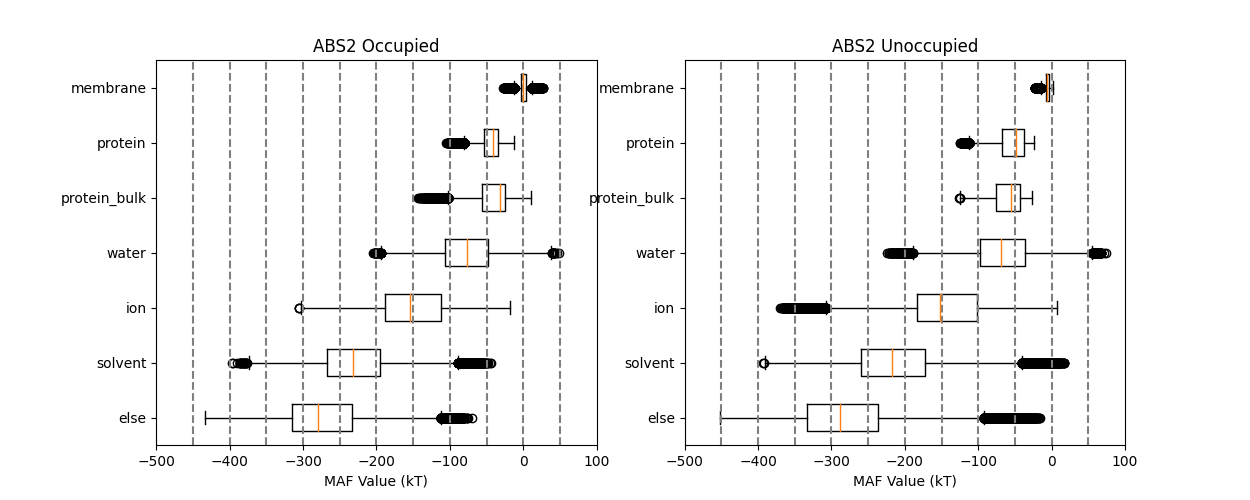}
\caption{Boxplots for the molecular ambiguity function (MAF) interactions of ATP with various system components for each of the two experiments.}
\label{fig:mcmaf_whisker}
\end{figure}

\subsection{Root Mean Square Deviation and Fluctuation}

\begin{figure}[h!]
\centering
\includegraphics[width=0.9\textwidth]{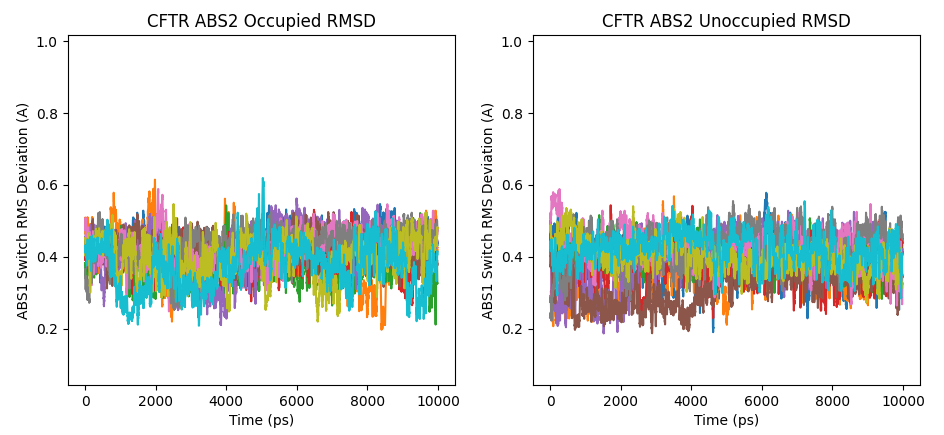}
\caption{The root-mean-square deviation for the Switch motif near both ATP binding sites over 10 trajectories for each of the two experiments. }
\label{fig:rmsd_d_loop}
\end{figure}

\begin{figure}[h!]
\centering
\includegraphics[width=0.9\textwidth]{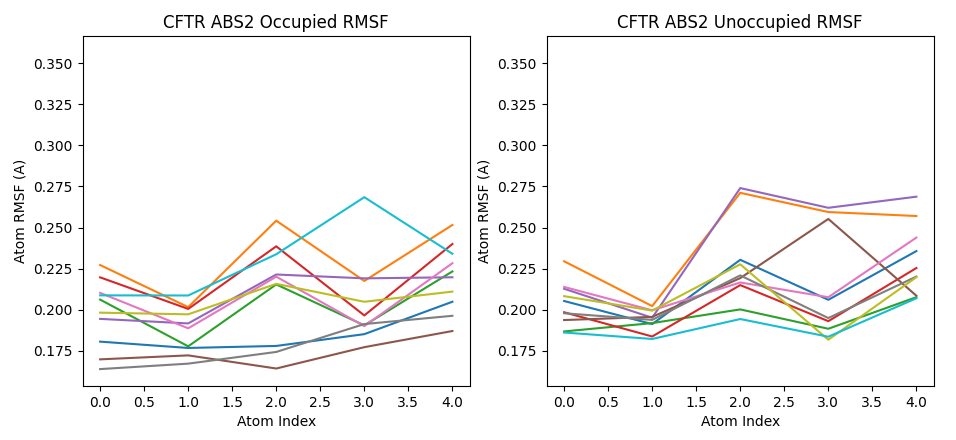}
\caption{The root-mean-square fluctuation for the Switch motif alpha carbons near both ATP binding sites over 10 trajectories for each of the two experiments. }
\label{fig:rmsf_d_loop}
\end{figure}

For the sake of verifying the molecular dynamics simulations for both experiments, we computed the root square deviation (RMSD) and root mean square fluctuation (RMSF) for all ten production runs for each experiment. The RMSD and RMSF were computed for the CFTR protein as a whole in addition to the six binding motifs around each of the ATP binding sites. The RMSD and RMSF for the D-loop motif is shown in Figures \ref{fig:rmsd_d_loop} and \ref{fig:rmsf_d_loop}, respectively. For both experiments, RMSF for all surrounding ABS2 motifs remained under 1~\AA, indicating that equilibration after ATP removal was successful. Figures for the RMSD and RMSF for the full CFTR protein and for other motifs surrounding ABS1 and ABS2 are available but not reported here.

\section{Conclusion}

In this work we have investigated interactions between the cystic fibrosis transmembrane conductance regulator protein and ATP. We applied molecular dynamics with randomly initialized experiments to study electrostatic attraction between ATP and CFTR. Electrostatic interactions were modeled using the previously developed molecular ambiguity function (MAF). We considered two cases -- one in which the competent ATP binding site of CFTR was already occupied by an ATP molecule and one in which it was unoccupied. We have demonstrated that, when CFTR is unbound from ATP, there is a favorable electrostatic potential gradient in the CFTR-ATP interaction that draws ATP towards the protein. We also compared the CFTR-ATP interaction to the membrane-ATP and solvent-ATP interactions to further contextualize our results. The electrostatic gradient is of comparable magnitude to solvent-ATP interactions, and the net electrostatic potential on ATP is mostly noise combined with the gradient towards CFTR.


\section{Acknowledgements}
The authors would like to thank Prof. James C. Gumbart for review of this document. \\
The authors would like to thank Dr. Michael A. Fiddy for sponsoring this work through the RadioBio program of the Defense Sciences Office (DSO): This material is based upon work supported by the Defense Advanced Research Projects Agency (DARPA) under Contract No. HR001117C0124. Any opinions, findings and conclusions or recommendations expressed in this material are those of the author(s) and do not necessarily reflect the views of the DARPA.



\bibliographystyle{ieeetr}
\bibliography{references}

\begin{thebibliography}{10}

\bibitem{maf}
K.~W. Allen, W.~D. Hunt, J.~D. Andreasen, J.~D. Farnum, A.~Saad-Falcon, R.~S.
  Westafer, and D.~R. Denison, ``Rigorous {Approach} to {Simulate}
  {Electromagnetic} {Interactions} in {Biological} {Systems},'' in {\em
  {NAECON} 2018 - {IEEE} {National} {Aerospace} and {Electronics}
  {Conference}}, (Dayton, OH), pp.~491--495, IEEE, July 2018.

\bibitem{katchalski-katzir_molecular_1992}
E.~Katchalski-Katzir, I.~Shariv, M.~Eisenstein, A.~A. Friesem, C.~Aflalo, and
  I.~A. Vakser, ``Molecular surface recognition: determination of geometric fit
  between proteins and their ligands by correlation techniques.,'' {\em
  Proceedings of the National Academy of Sciences}, vol.~89, no.~6,
  pp.~2195--2199, 1992.

\bibitem{gabb_modelling_1997}
H.~A. e.~a. Gabb, ``Modelling protein docking using shape complementarity,
  electrostatics and biochemical information,'' {\em Journal of Molecular
  Biology}, p.~15, 1997.

\bibitem{gumbart_efficient_2013}
J.~C. Gumbart, B.~Roux, and C.~Chipot, ``Efficient {Determination} of
  {Protein}–{Protein} {Standard} {Binding} {Free} {Energies} from {First}
  {Principles},'' {\em Journal of Chemical Theory and Computation}, vol.~9,
  pp.~3789--3798, Aug. 2013.

\bibitem{csanady_structure_2019}
L.~Csanády, P.~Vergani, and D.~C. Gadsby, ``Structure, {Gating}, and
  {Regulation} of the {CFTR} {Anion} {Channel},'' {\em Physiological Reviews},
  vol.~99, pp.~707--738, Jan. 2019.

\bibitem{srikant_evolutionary_2020}
S.~Srikant, ``Evolutionary history of {ATP}‐binding cassette proteins,'' {\em
  FEBS Letters}, vol.~594, pp.~3882--3897, Dec. 2020.

\bibitem{cui_electrophysiological_2021}
G.~Cui, K.~A. Cottrill, and N.~A. McCarty, ``Electrophysiological {Approaches}
  for the {Study} of {Ion} {Channel} {Function},'' in {\em Structure and
  {Function} of {Membrane} {Proteins}} (I.~Schmidt-Krey and J.~C. Gumbart,
  eds.), pp.~49--67, New York, NY: Springer US, 2021.

\bibitem{namd}
J.~C. Phillips, D.~J. Hardy, J.~D.~C. Maia, J.~E. Stone, J.~V. Ribeiro, R.~C.
  Bernardi, R.~Buch, G.~Fiorin, J.~Hénin, W.~Jiang, R.~McGreevy, M.~C.~R.
  Melo, B.~K. Radak, R.~D. Skeel, A.~Singharoy, Y.~Wang, B.~Roux,
  A.~Aksimentiev, Z.~Luthey-Schulten, L.~V. Kalé, K.~Schulten, C.~Chipot, and
  E.~Tajkhorshid, ``Scalable molecular dynamics on {CPU} and {GPU}
  architectures with {NAMD},'' {\em The Journal of Chemical Physics}, vol.~153,
  p.~044130, July 2020.

\bibitem{pdb}
H.~M. Berman, ``The {Protein} {Data} {Bank},'' {\em Nucleic Acids Research},
  vol.~28, pp.~235--242, Jan. 2000.

\bibitem{6msm}
Z.~Zhang, F.~Liu, and J.~Chen, ``Molecular structure of the {ATP}-bound,
  phosphorylated human {CFTR},'' {\em Proceedings of the National Academy of
  Sciences}, vol.~115, pp.~12757--12762, Dec. 2018.

\bibitem{cftr_chol}
G.~Cui, K.~A. Cottrill, K.~M. Strickland, S.~A. Mashburn, M.~Koval, and N.~A.
  McCarty, ``Alteration of {Membrane} {Cholesterol} {Content} {Plays} a {Key}
  {Role} in {Regulation} of {Cystic} {Fibrosis} {Transmembrane} {Conductance}
  {Regulator} {Channel} {Activity},'' {\em Frontiers in Physiology}, vol.~12,
  p.~652513, June 2021.

\bibitem{charmm}
K.~Vanommeslaeghe, E.~Hatcher, C.~Acharya, S.~Kundu, S.~Zhong, J.~Shim,
  E.~Darian, O.~Guvench, P.~Lopes, I.~Vorobyov, and A.~D. Mackerell, ``{CHARMM}
  general force field: {A} force field for drug-like molecules compatible with
  the {CHARMM} all-atom additive biological force fields,'' {\em Journal of
  Computational Chemistry}, pp.~NA--NA, 2009.

\bibitem{namd_potentials}
``{Potential energy functions}.''
  \url{https://www.ks.uiuc.edu/Research/namd/2.10/ug/node22.html}.
\newblock [Online; accessed 26-July-2021].

\bibitem{debye}
D.~Lockhart and P.~Kim, ``Electrostatic screening of charge and dipole
  interactions with the helix backbone,'' {\em Science}, vol.~260,
  pp.~198--202, Apr. 1993.

\bibitem{maxwellfreq}
J.~R. de~Xammer~Oro, G.~Ruderman, J.~R. Grigera, and F.~Vericat, ``Threshold
  frequency for the ionic screening of electric fields in electrolyte
  solutions,'' {\em Journal of the Chemical Society, Faraday Transactions},
  vol.~88, no.~5, p.~699, 1992.

\bibitem{psi4}
D.~G.~A. Smith, L.~A. Burns, A.~C. Simmonett, R.~M. Parrish, M.~C. Schieber,
  R.~Galvelis, P.~Kraus, H.~Kruse, R.~Di~Remigio, A.~Alenaizan, A.~M. James,
  S.~Lehtola, J.~P. Misiewicz, M.~Scheurer, R.~A. Shaw, J.~B. Schriber, Y.~Xie,
  Z.~L. Glick, D.~A. Sirianni, J.~S. O’Brien, J.~M. Waldrop, A.~Kumar, E.~G.
  Hohenstein, B.~P. Pritchard, B.~R. Brooks, H.~F. Schaefer, A.~Y. Sokolov,
  K.~Patkowski, A.~E. DePrince, U.~Bozkaya, R.~A. King, F.~A. Evangelista,
  J.~M. Turney, T.~D. Crawford, and C.~D. Sherrill, ``Psi4 1.4: {Open}-source
  software for high-throughput quantum chemistry,'' {\em The Journal of
  Chemical Physics}, vol.~152, p.~184108, May 2020.

\bibitem{pme}
T.~Darden, D.~York, and L.~Pedersen, ``Particle mesh {Ewald}: {An} {N}log({N})
  method for {Ewald} sums in large systems,'' {\em The Journal of Chemical
  Physics}, vol.~98, pp.~10089--10092, June 1993.

\end{thebibliography}
\nocite{*}
\end{document}